\documentstyle[times,epsfig]{jaa}

\begin{document}

\title[Spectral variability of FSRQs]
{The spectral variability of FSRQs}
\author[M. F. Gu \& Y. L. Ai]
       {Minfeng Gu$^{1}$\thanks{e-mail:gumf@shao.ac.cn} \& Y. L. Ai$^{2,3}$ \\
        $^{1}$ Key Laboratory for Research in Galaxies and Cosmology, Shanghai Astronomical \\
        Observatory,
    Chinese Academy of Sciences, 80 Nandan Road, Shanghai 200030, China\\
    $^{2}$ National Astronomical Observatories/Yunnan Observatory, Chinese Academy of \\
    Sciences, Kunming, Yunnan, P.O. BOX 110, China\\
$^{3}$ Key Laboratory for the Structure and Evolution of Celestial
Objects, Chinese Academy\\
 of Sciences, Kunming, China}
\maketitle
\label{firstpage}

\begin{abstract}
The optical variability of 29 flat spectrum radio quasars in SDSS
Stripe 82 region are investigated by using DR7 released multi-epoch
data. All FSRQs show variations with overall amplitude ranging from
0.24 mag to 3.46 mag in different sources. About half of FSRQs show
a bluer-when-brighter trend, which is commonly observed for blazars.
However, only one source shows a redder-when-brighter trend, which
implies it is rare in FSRQs. In this source, the thermal emission
may likely be responsible for the spectral behavior.
\end{abstract}

\begin{keywords}
galaxies: active -- galaxies: quasars: general -- galaxies:
photometry
\end{keywords}

\section{Introduction}

Blazars, including BL Lac objects and flat-spectrum radio quasars
(FSRQs), are the most extreme class of active galactic nuclei
(AGNs), characterized by strong and rapid variability, high
polarization, and apparent superluminal motion. These extreme
properties are generally interpreted as a consequence of non-thermal
emission from a relativistic jet oriented close to the line of
sight. While the jet non-thermal emission are thought to be
dominated in FSRQs, this is actually not always true. As a matter of
fact, Chen, Gu \& Cao (2009) found that the thermal emission can be
dominant at least in optical bands for some FSRQs.

The color behaviors in blazars are still in debates. The
bluer-when-brighter trend (BWB) is commonly observed in blazars
(e.g. Fan et al. 1998; Raiteri et al. 2001; Villata et al. 2002; Wu
et al. 2007). But, opposite examples have also been found (e.g. Gu
et al. 2006; Dai et al. 2009; Rani et al. 2010), such as 3C 454.3
(Gu et al. 2006), PKS 0736+017 (Clements et al. 2003; Ram\'{i}rez et
al. 2004), 3C 446 (Miller 1981), PKS 1622-297 \& CTA 102 (Osterman
Meyer et al. 2008, 2009). But still not many FSRQs were found to
show redder-when-brighter trend (RWB). From a sample of FSRQs
selected from SDSS, we briefly shown here one more FSRQ with RWB
(see Gu \& Ai 2011 for details).

\section{Sample}

The Stripe-82 region, i.e. right ascension $\alpha = 20^{\rm h} -
4^{\rm h}$ and declination $\delta=-1^{\circ}.25 - +1^{\circ}.25$,
was repeatedly scanned during the SDSS-I phase (2000 - 2005) and
also over the course of three 3-month campaigns in three successive
years in 2005 - 2007 known as the SDSS Supernova Survey. Those
quasars selected from SDSS DR7 quasars catalogue (Schneider et al.
2010) in the region of the Stripe 82, were cross-correlated with the
Faint Images of the Radio Sky at Twenty Centimeters (FIRST) 1.4-GHz
radio catalogue (Becker, White \& Helfand 1995), the Green Bank 6-cm
(GB6) survey at 4.85 GHz radio catalogue (Gregory et al. 1996), and
the Parkes-MIT-NRAO (PMN) radio continuum survey at 4.85 GHz
(Griffith \& Wright, 1993). A quasar is defined as a FSRQ according
to the radio spectral index between 1.4 and 4.85 GHz with
$\alpha<0.5$ ($f_{\nu}\propto\nu^{-\alpha}$), which resultes in a
sample of 32 FSRQs.

\section{Results}

We use the photometric data obtained during the SDSS-I phase from
Data Release 7 (DR7; Abazajian et al. 2009) and the SN survey during
2005 - 2007. We use the point-spread function magnitudes. The
spectral index $\alpha$ are calculated from the linear fit on $\rm
log~\it f_{\nu} - \rm log ~\nu$ after extinction correction on
$ugriz$ flux density.

\subsection{Variability}

Among 32 FSRQs, three sources were excluded in our analysis due to
various reasons. All remaining 29 FSRQs show large amplitude
variations with overall variations in r band $\Delta r=0.24 - 3.46$
mag, which is much larger than that of radio quiet AGNs, 0.05 - 0.3
mag (e.g. Ai et al. 2010), however is typical for blazars (e.g. Gu
et al. 2006). There are four sources with $\Delta r>1$ mag, i.e.
SDSS J001130.400+005751.80 - $\Delta r=3.46$ mag; SDSS
J023105.597+000843.61 - $\Delta r=1.02$ mag; SDSS
J025515.096+003740.55 - $\Delta r=1.70$ mag; SDSS
J235936.817-003112.78 - $\Delta r=1.20$ mag. In general, the
variations in different bands show similar trends.

\subsection{Spectral index \& brightness relation}

The correlation between the spectral index $\alpha_{\nu}$ and psf
$r$ magnitude were checked for all sources using the Spearman rank
correlation analysis method. We found that 15 of 29 FSRQs show a
significant correlation at a confidence level of $>99\%$, of which
14 FSRQs show positive correlations, and only one FSRQ (SDSS
J001130.40$+$005751.7) shows a negative correlation.

\begin{figure}
\centering
\includegraphics[width=0.8\textwidth]{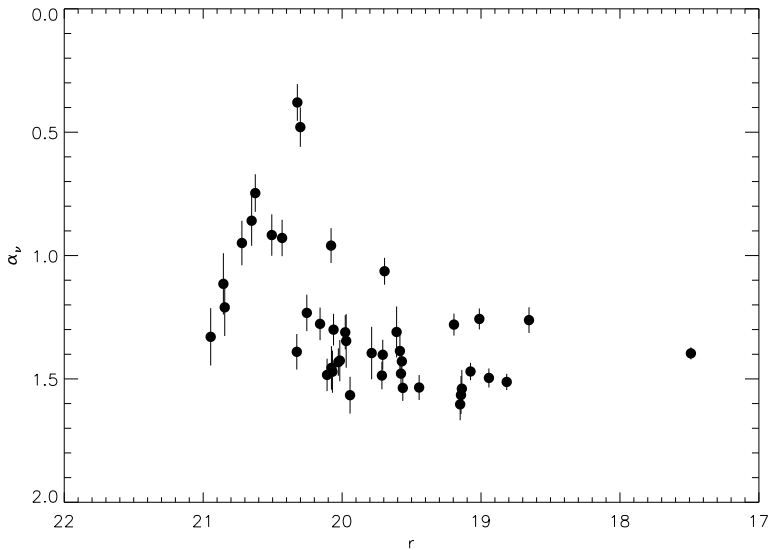}
\caption{The relationship between the spectral index and the psf
magnitude at r band for SDSS J001130.40$+$005751.7. A significant
anti-correlation is present, which implies a redder-when-brighter
trend.} \label{fig1}
\end{figure}

The negative correlation shows the source becomes steeper when
source is brighter, i.e. RWB, which is shown in Fig. 1 with total 43
datapoints. The Spearman correlation analysis shows a significant
negative correlation with a correlation coefficient of $r_{\rm
s}=-0.606$ at confidence level of $>99.99\%$. This source is
included in the first Fermi Large Area Telescope AGNs catalogue with
photon index of $2.51\pm0.15$ and it is classified as a
low-synchrotron-peaked FSRQs with $\nu_{\rm peak}<10^{14}$ Hz (Abdo
et al. 2010). With the source redshift $z=1.4934$, SDSS $ugriz$
wavebands corresponds to the wavelength range of $1424 - 3582 \AA$
in the source rest frame, which therefore is likely at the falling
part of synchrotron SED. Shang et al. (2005) show that the spectral
break happens at around $1100~ \AA$, which is thought to be closely
related with big blue bump. If this spectral break also exists in
SDSS J001130.40$+$005751.7, we would expect to observe the rising
part of accretion disk thermal emission when it dominates over the
nonthermal emission. Indeed, Fig. 1 qualitatively shows that the
optical spectral is rising ($\alpha_{\nu}<1.0$) when the source is
at low flux state, while it becomes falling ($\alpha_{\nu}>1.0$)
when the source is at high flux state implying a nonthermal low peak
frequency synchrotron emission starts to be dominated.

It is interesting to note that FSRQs are supposed to generally have
redder-when-brighter trend (e.g. Dai et al. 2009), for example, in
at least two of three, 3C 454.3 and PKS 0420-014 (Gu et al. 2006),
and four of the six, PKS 0420-014, 4C 29.45, PKS 1510-089 and 3C
454.3 (Rani et al. 2010), all of which are known
low-synchrotron-peaked FSRQs. However, our results imply that the
redder-when-brighter trend may likely be rare in FSRQs, at least for
our present sample. Although the details is unclear, it is most
likely that the spectral behaviors of FSRQs is dependent on the
position of synchrotron peak frequency, the sampled optical
wavelength range in the source rest frame, the positions of thermal
blue bump and its strength compared to jet emission (see also Gu \&
Ai 2011).

\section{Conclusions}

For a sample of 29 FSRQs selected in SDSS stripe 82 region, all
FSRQs show large amplitude overall variations, e.g. 0.24 to 3.46 mag
at r band. We only found the significant negative correlation
between the spectral index and r magnitude (i.e. RWB) in one FSRQ.
This implies RWB trend is rare in FSRQs, which could be explained by
the contribution of thermal accretion disk emission. In contrast,
BWB is more common in FSRQs.

This work is supported by the National Science Foundation of China
(grants 10703009, 10821302, 10833002, 11033007 and 11073039), and by
the 973 Program (No. 2009CB824800).


\label{lastpage}

\end{document}